\documentclass[floats,aps,showpacs,twocolumn]{revtex4}

\usepackage[dvips]{graphicx}
\usepackage{amsmath}   
\usepackage{amsfonts}

\newcommand{\heff}{h_{\text{eff}}}
\newcommand{\heffbar}{\hbar_{\text{eff}}}
\newcommand{\Areg}{A_{\text{reg}}}
\newcommand{\Dreg}{\Delta_{\text{reg}}}
\newcommand{\Dcha}{\Delta_{\text{ch}}}
\newcommand{\Ncha}{N_{\text{ch}}}
\newcommand{\Nreg}{N_{\text{reg}}}
\newcommand{\Ntot}{N_{\text{tot}}}

\newcommand{\freg}{f_{\text{reg}}}
\newcommand{\cFreg}{{\cal F}_{\text{reg}}}
\newcommand{\ffl}{f_{\text{{f}l}}}
\newcommand{\tH}{\tau_{\text{\tiny{H,ch}}}}

\newcommand{\mmax}{m_{\text{max}}}
\newcommand{\mstar}{m^{*}}
\newcommand{\mfl}{\mstar}

\newcommand{\PSImagx}[2]{\includegraphics[width=#2]{#1}}

\newcommand{\N}{\mathbb{N}}

\newcommand{\ud}{\text{d}}
\newcommand{\ue}{\text{e}}
\newcommand{\ui}{\text{i}}

\newcommand{\emm}{m}  
\newcommand{\Wnull}{W_0}
\newcommand{\Weins}{W_1}
\newcommand{\Wm}{W_\emm}
\newcommand{\Weq}{W_{\text{eq}}}

\begin{document}

\title{Universality in the flooding of regular islands by chaotic states}

\author{Arnd B\"acker$^1$, Roland Ketzmerick$^1$, 
and Alejandro G. Monastra$^{1,2}$}

\affiliation{$^1$Institut f\"ur Theoretische Physik, Technische
Universit\"at Dresden, 01062 Dresden, Germany\\
$^2$Departamento de F\'{\i}sica, Comisi\'on Nacional de Energ\'{\i}a At\'omica, 
Av. del Libertador 8250, 1429 Buenos Aires, Argentina}

\date{05.01.2007, revised 03.05.2007}

\begin{abstract}

We investigate the structure of eigenstates in 
systems with a mixed phase space
in terms of their projection onto individual regular tori.
Depending on dynamical tunneling rates and the Heisenberg time,
regular states disappear and chaotic states flood the regular tori.
For a quantitative understanding
we introduce a random matrix model.
The resulting statistical properties of
eigenstates as a function of an effective coupling strength
are in very good agreement with numerical results for a kicked system.
We discuss the implications of these results for the applicability
of the semiclassical eigenfunction hypothesis.
\end{abstract}

\maketitle

\section{Introduction}

The classical dynamics in Hamiltonian systems shows a rich behaviour
ranging from integrable to fully chaotic motion. In chaotic systems
nearby trajectories separate exponentially in time and ergodicity
implies that a typical trajectory fills out the energy-surface in a
uniform way. However, integrable and fully chaotic dynamics are
exceptional \cite{MarMey74} as typical Hamiltonian systems show a
mixed phase space in which regions of regular motion, the
so-called regular islands around stable periodic orbits, and
chaotic dynamics, the so-called chaotic sea, coexist.

For quantized Hamiltonian systems the fundamental questions concern
the behaviour of the eigenvalues and the properties of eigenfunctions,
especially in the semiclassical regime.
From the semiclassical eigenfunction hypothesis
\cite{Per73,Vor76,Ber77b,Vor79,Ber83} one expects that in the
semiclassical limit the eigenstates concentrate on those regions in
phase space which a typical orbit explores in the long-time limit. For
integrable systems these are the invariant tori.
In contrast, for ergodic systems almost all orbits fill
the energy shell in a uniform way. For this situation
the semiclassical eigenfunction hypothesis 
is proven by the quantum ergodicity theorem which shows
that almost all eigenstates become equidistributed on the energy 
shell~\cite{Qerg}.

For systems with a mixed phase space, in the semiclassical limit $(h
\rightarrow 0)$, the semiclassical eigenfunction hypothesis implies
that the eigenstates can be classified as being either regular or
chaotic according to the phase-space region on which they
concentrate. This is supported by several studies, see
e.g.~\cite{BohTomUll93,ProRob93b,LiRob95b,CarVerFen98,VebRobLiu99,MarKee2005}.
It is also possible, that the influence of a
regular island  quantum mechanically extends beyond the outermost
invariant curve due to partial barriers like cantori 
and that quantization conditions remain 
approximately applicable even outside of the island \cite{BohTomUll93}.
However, it was recently shown that 
the classification into regular and chaotic states
does not hold when the phase
space has an infinite volume \cite{HufKetOttSch2002}. In this case
eigenstates may completely ignore the classical phase space boundaries
between regular and chaotic regions.

In order to understand the behaviour of eigenstates away from the
semiclassical limit, i.e.\ at finite values of the Planck constant
$h$, one has to compare the size of phase-space structures with $h$.
Let us consider for simplicity the case of two-dimensional area
preserving maps and their quantizations. Regular states of an island
concentrate on tori which fulfill the EBK-type
quantization condition
\begin{equation} \label{eq:EBK}
 \oint p \, \ud q = (m+1/2) h \qquad m=0, 1, ...
\end{equation}
for the enclosed area \cite{BerBalTabVor79}. This quantization rule
explicitly shows that regular eigenstates only appear if $h/2$ is
smaller than the area $\Areg$ of that island.

Another consequence of finite $h$ in systems with a mixed phase space
is dynamical tunneling
\cite{DavHel81}, i.e.\ tunneling through dynamically generated
barriers in phase space, in contrast to the usual tunneling under a
potential barrier. Dynamical tunneling couples the subspace spanned by
the regular basis states, corresponding to the quantization condition
\eqref{eq:EBK}, with the complementary subspace \cite{ftn2}
composed of chaotic
basis states. This raises the question whether the {\it eigenstates}
of such a quantum system
can still be called {\it regular} or {\it chaotic}.

In Ref.~\cite{BaeKetMon2005} it was shown that \eqref{eq:EBK} is not a
sufficient condition for the existence of a regular eigenstate on the
$m$-th quantized torus. In addition one has to fulfill
\begin{equation}
  \gamma_m < \frac{1}{\tH} , \label{newcondition}
\end{equation}
where $\tH = h / \Dcha$ is the Heisenberg time of the surrounding
chaotic sea with mean level spacing $\Dcha$ and $\gamma_m$ is the
decay rate of the $m$-th regular state, if the chaotic sea were
infinite.
When condition \eqref{newcondition} is violated one observes
eigenstates which extend over the chaotic region and flood the $m$-th
torus \cite{BaeKetMon2005}. To distinguish them from the chaotic
eigenstates that do not flood the torus, they are referred to as 
{\it flooding eigenstates}. For the limiting case of complete flooding of
all tori, the corresponding eigenstates were called amphibious
\cite{HufKetOttSch2002}. Recently, the consequences of flooding for the
transport properties in rough nano-wires were 
studied \cite{FeiBaeKetRotHucBur2006}.

The process of flooding was explained and
demonstrated for a kicked system 
in Ref.~\cite{BaeKetMon2005}.
Condition \eqref{newcondition} was
obtained by scaling arguments, which cannot provide a
prefactor. Moreover, for an ensemble of systems, one would like to know
the probability for the existence of a regular
eigenstate. In particular, when varying the Heisenberg time, how broad
is the transition regime during which this probability goes from 1 to 0? 
Another question is, how do the chaotic eigenstates turn into flooding
eigenstates for a given torus?

In this paper we give quantitative answers to these questions. We
study the flooding of regular tori in terms of the weight of
eigenstates inside the regular region and devise a random matrix model
which allows for describing the statistics of these weights in
detail. Random matrix models have been very successful for obtaining
quantitative predictions on eigenstates in both fully chaotic systems
and systems with a mixed phase space, see
e.g.~\cite{KusMosHaa1988,BohTomUll93,HaaZyc1990,TomUll1994,Bae2003,KeaMezMon2003}.
For the present situation we propose a random matrix model which takes
regular basis states and their coupling to the chaotic basis states
into account. The only free parameters are the strength of the
coupling and the ratio of the number of regular to the number
of chaotic basis states. From this
model the weight distribution for eigenstates is determined.

For a kicked system we define the weight by the projection of the
eigenstates onto regular basis states localized on a given torus $m$. The
distribution of the weights allows for studying the flooding of each torus
separately. The resulting distributions are compared with the
prediction of the random matrix model and, after an appropriate
rescaling, very good agreement is observed. This agreement shows
explicitly the universal features underlying the process of flooding,
giving a precise criterion for the existence or non--existence of
regular, chaotic, and flooding eigenstates in mixed systems.

The text is organized as follows. In section \ref{TheSystem} we
introduce the kicked system used for the numerical illustrations, both
classically (part A) and quantum mechanically (part B). In section
\ref{TheSystem}~C we define the weight of an eigenstate by its projection
onto regular basis states and investigate the distribution of the
weights for the kicked system. In section \ref{RMModel} we 
introduce the random matrix model
and determine the corresponding weight distribution as a function of
the coupling strength. In section \ref{Comparison} the relation
between parameters of the kicked system and the random matrix model is
derived. This allows for a direct comparison of the distributions. In
section \ref{Fraction} we consider the fraction of regular
eigenstates, both for an individual torus and for the entire island. 
In section \ref{Fraction-flood} we briefly discuss  
the consequences of the random matrix model on the 
number of flooding eigenstates. 
A summary and discussion of the eigenfunction
structure in generic systems with a mixed phase space
is given in section \ref{Conclusions}.

\section{The Kicked System} \label{TheSystem}

\subsection{Classical dynamics}

\begin{figure}[b]
  \begin{center}
    \PSImagx{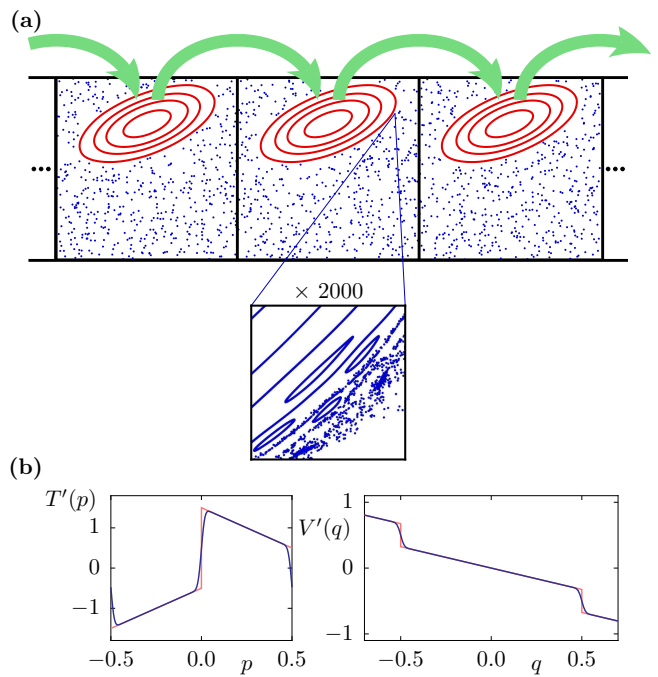}{8.6cm}
    \caption{(color online) 
              (a) Classical dynamics of the kicked system given by 
              Eqs.~(\ref{mapQ}) and (\ref{mapP}). 
              Invariant tori of the regular island are
              shown (continuous curves) and the transport to the right is
              indicated by the arrows. The dots correspond to one
              chaotic orbit. The magnification shows that the
              boundary of the island to the chaotic sea is rather
              sharp with only very small secondary islands.
              (b) Smoothed functions $T'(p)$ and $V'(q)$ (blue, dark lines)
              and discontinuous functions $t'(p)$ and $v'(q)$ (red, light
              lines) according to Eqs.~(\ref{tp}-\ref{Vq}).
     \label{fig:system}} 
   \end{center}
\end{figure}

For a general one-dimensional kicked Hamiltonian
\begin{equation}
  H(p,q,t) =  T(p) + V(q) \sum_{n= -\infty}^{\infty} \delta(t - n ) \ ,
   \label{hamiltonian}
\end{equation}
the dynamics is fully determined by the mapping of position and
momentum $(q_n,p_n)$ at times $t=n + 0^+$ just after the kicks
\begin{eqnarray}
  q_{n+1} &=& q_n + T'(p_n) \ , \label{mapQ} \\
  p_{n+1} &=& p_n - V'(q_{n+1}) \ . \label{mapP}
\label{mapping}
\end{eqnarray}
Choosing the functions $T'(p)$ and $V'(q)$ appropriately, one can
obtain a system with a large regular island and a homogeneous chaotic
sea.  For the system considered in \cite{BaeKetMon2005}, first
introduced in \cite{HufKetOttSch2002}, one starts with
the piecewise linear functions (see Fig.~\ref{fig:system}b)
\begin{eqnarray}
  t'(p) &=& \frac{1}{2} + \left( \frac{1}{2} - s p \right) {\rm sign}
  \left( p - \lfloor p + 1/2 \rfloor \right) \ , \label{tp} \\
  v'(q) &=& -r q - (1-r) \lfloor q + 1/2 \rfloor \ , \label{vq}
\end{eqnarray}
where $\lfloor x \rfloor$ is the floor function, and $s$ and $r$ are
two parameters determining the properties of the regular island
and the chaotic sea. 
Using a Gaussian smoothing with $G_{\varepsilon}
(z)=\exp(-z^2/2\varepsilon^2)/ \sqrt{2\pi\varepsilon^2}$, one obtains
analytic functions
\begin{eqnarray}\label{functionsprime}
  T'(p) &=& \int_{-\infty}^{\infty} \ud z \; t'(p+z) \; G_{\varepsilon}(z) \ ,
            \label{Tp} \\
  V'(q) &=& \int_{-\infty}^{\infty} \ud z \; v'(q+z) \; G_{\varepsilon}(z)  \ .
            \label{Vq}
\end{eqnarray}
By construction, these functions have the periodicity properties
\begin{eqnarray}\label{periodicityprime}
  T'(p+k) &=& T'(p)   \ , \\
  V'(q+k) &=& V'(q)-k  \ ,\label{periodicityprime2}
\end{eqnarray}
for any integer $k$. We consider $p\in[-1/2, 1/2[$ and
$q\in[-1/2,-1/2+M[\,$ with periodic boundary conditions. The phase space
is composed of a chain of transporting islands centered at $(\bar
q,\bar p)=(k,1/4)$ with $0 \leq k \leq M-1$ that are mapped one unit
cell to the right (see Fig.~\ref{fig:system}a). The surrounding chaotic
sea has an average drift to the left as the overall transport is
zero \cite{SchOttKetDit2001,SchDitKet2005}. 
The fine scale structure at the boundary of the island to the
chaotic sea has a very small area (see the magnification in
Fig.~\ref{fig:system}a). Resonances in this layer are irrelevant in
the $h$ regime studied here.
For $s=2$, $r=0.65$
and $\varepsilon=0.015$ the regular island has a relative area $\Areg
\approx 0.215$.

\subsection{Quantization} \label{Quantization}

In kicked systems, the quantum evolution of a state after one period of time
\begin{equation}
  |\psi (t + 1) \rangle = \hat{U} |\psi (t) \rangle \ , \label{evolution}
\end{equation}
is fully determined by the unitary operator, see e.g.\
\cite{BerBalTabVor79,HanBer80,ChaShi86,Esp93,DegGra03},
\begin{equation}
  \hat{U}= \exp\left( -\frac{2 \pi \ui}{\heff} V( \hat{q} ) \right) 
           \exp\left( -\frac{2 \pi \ui}{\heff} T( \hat{p} ) \right) 
                             \ . \label{propagator}
\end{equation}
Here the effective Planck's constant $\heff$ is Planck's constant $h$
divided by the size of one unit cell. The eigenstates of this operator
are defined by
\begin{equation} \label{eq:eigenstates}
  \hat{U} |\psi_j \rangle = \ue^{2 \pi \ui \varphi_j} |\psi_j \rangle  \ , 
\end{equation}
where the eigenphase $\varphi_j$ is the quasienergy divided by 
$\hbar\omega$. 
In order to fulfill the periodicity of the classical dynamics
in $p$ direction, the quantum states have to obey the
quasi-periodicity condition
\begin{equation}
  \langle p + 1 |\psi \rangle 
    = \ue^{- 2\pi\ui \chi_p} \langle p  |\psi \rangle  \ .
  \label{quantumperiodP}
\end{equation}
One can show that this leads to quantum states that are a linear
combination of the discretized position states $| q_j \rangle$, with
$q_j = \heff ( j + \chi_p)$. Additionally, imposing periodicity after
$M$ unit cells in $q$ direction, quantum states have to fulfill the
property
\begin{equation}
  \langle q + M |\psi \rangle = 
     \ue^{2\pi\ui \chi_q} \langle q  |\psi \rangle  \ . \label{quantumperiodQ}
\end{equation}
Because of the required periodicity the phase space is compact and the
effective Planck's constant can only be a rational number
\begin{equation}
  \heff = \frac{M}{N}   \ . \label{quanthbar}
\end{equation}
We consider the case of incommensurate $M$ and $N$, so that the
quantum system is not effectively reduced to less than $M$ cells.

\begin{figure}[t]
  \begin{center}
    \PSImagx{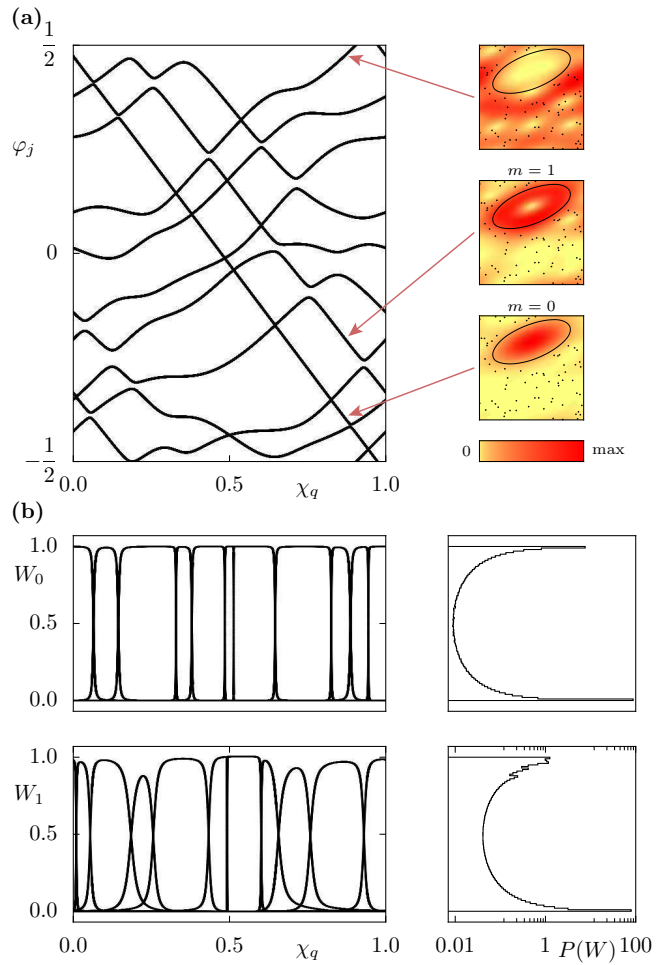}{8.6cm}

    \caption{(color online) (a) Eigenphases of the kicked system vs
              $\chi_q$ for $\heff=1/10$. The pattern of straight lines
              (interrupted by avoided crossings) with negative slope
              corresponds to regular eigenstates with $m=0$ and $m=1$
              whose Husimi functions are shown to the right. The other
              eigenstates are chaotic and live outside of the
              regular region, as can be seen from the Husimi
              representation. (b) Weights $\Wnull$ and $\Weins$ of all
              eigenstates vs $\chi_q$ (left). Distribution $P(W)$ of
              these weights in a log-linear representation (right).
              }
              \label{fig:Avoided} 
  \end{center}
\end{figure}

The properties \eqref{periodicityprime}, \eqref{periodicityprime2}
of $T'(p)$ and $V'(q)$ imply for their integrals
\begin{eqnarray}
T(p+k) &=& T(p)   \ , \label{periodicityT} \\
V(q+k) &=& V(q)- k q - \frac{k^2}{2} \ . \label{periodicityV}
\end{eqnarray}
From this one finds that the propagator $\hat U$ is consistent with
the periodicity conditions \eqref{quantumperiodP} and
\eqref{quantumperiodQ} if and only if
\begin{equation}
  M \left( \chi_p + \frac{N}{2} \right)  \in \mathbb{Z}  \ . 
  \label{quantcondition}
\end{equation}
For given $M$ and $N$, this condition limits the possible values of
the phase $\chi_p$, while $\chi_q$ remains arbitrary. Thus, in the
basis given by the position states $|q_j \rangle$, with $0 \leq j \leq
N-1$, where $N$ is the dimension of the Hilbert space, the propagator
$\hat U$ is represented by the finite $N \times N$ unitary matrix

\begin{equation} \label{Umatrix}
   U_{k l} = \frac{1}{N} \sum_{j=0}^{N-1} \ue^{ -\frac{\ui}{\heffbar} 
           \left[ V(q_k) + T(p_j) + p_j (q_l - q_k) \right]}    \ , 
\end{equation}
where $0 \leq k,l \leq N-1$ and $p_j = (j+\chi_q)/N$. Finding the
solution of \eqref{eq:eigenstates}, i.e. the eigenphases and
eigenstates of the system, therefore reduces to the numerical
diagonalization of the matrix \eqref{Umatrix}.
The result is illustrated in
Fig.~\ref{fig:Avoided}(a) for $\heff = 1/10$, where the eigenphases
are plotted as a function of $\chi_q$. The straight lines with
negative slope correspond to the regular eigenstates
\cite{SchOttKetDit2001,SchDitKet2005}, whose Husimi distributions are
shown to the right in Fig.~\ref{fig:Avoided}(a).  Lines with an
average positive slope correspond to chaotic eigenstates.

When the system consists of $M$ unit cells one has $M$ regular basis
states localized on the $m$-th torus. Their EBK eigenphases are
equispaced with a distance $1/M$ \cite{Sch:PhD-BaeKetMonSch}.

\subsection{Projection onto regular basis states} \label{Projection}

In order to investigate the amount of flooding we use the projection
of the eigenstates onto regular basis states of the island region.
For the considered kicked system regular basis states can be
constructed from harmonic oscillator eigenstates, as the invariant
tori are accurately approximated by ellipses
\cite{Sch:PhD-BaeKetMonSch}. The expression for
the $m$-th harmonic oscillator state, centered in a phase space point
$(\bar q,\bar p)$, is
\begin{align}
   \langle q &| \varphi^{\emm}_{\bar q,\bar p} \rangle =
   \frac{1}{\sqrt{2^m m!}} \left( \tfrac{{\rm Re}~\sigma}{\pi \heffbar}
   \right)^{1/4} H_m \left( \sqrt{\tfrac{{\rm Re}~\sigma}{\heffbar}} (q
   - \bar q) \right) \nonumber\\
    & \times \exp \left( - \tfrac{\sigma}{2 \heffbar} (q- \bar q)^2 +
    \tfrac{\ui}{\heffbar} \bar p (q - \bar q /2) \right) 
   \label{tiltedstate}
\end{align}
where $H_m$ is the Hermite polynomial of degree $m$. The complex
constant $\sigma$ takes into account the squeezing and rotation of the
state. From the linearized map at the stable fixed point of the island
one finds $\sigma = (\sqrt{351}- 13\, \ui )/40$.

For a chain with $M$ identical 
cells, a regular basis state is a linear combination of
the harmonic oscillator states $|\varphi^{\emm}_{k,1/4} \rangle$,
centered in the $k$-th island for $0 \leq k \leq M-1$ and properly
normalized and periodized in the $q$ and $p$ directions
\cite{Sch:PhD-BaeKetMonSch}. The subspace spanned by these $M$ regular basis
states is the same as the one spanned by the $M$ harmonic oscillator
states $|\varphi^{\emm}_{k,1/4} \rangle$.  Therefore, we define the
weight $\Wm$ of a normalized state $| \Psi \rangle$ by its projection
onto this subspace corresponding to the $m$-th quantized torus
\begin{equation}
  \Wm = \sum_{k=0}^{M-1} | \langle \Psi | \varphi^{\emm}_{k,1/4}
  \rangle |^2 \ . \label{weightsemiclass}
\end{equation}

\begin{figure}[b]
  \begin{center}
    \PSImagx{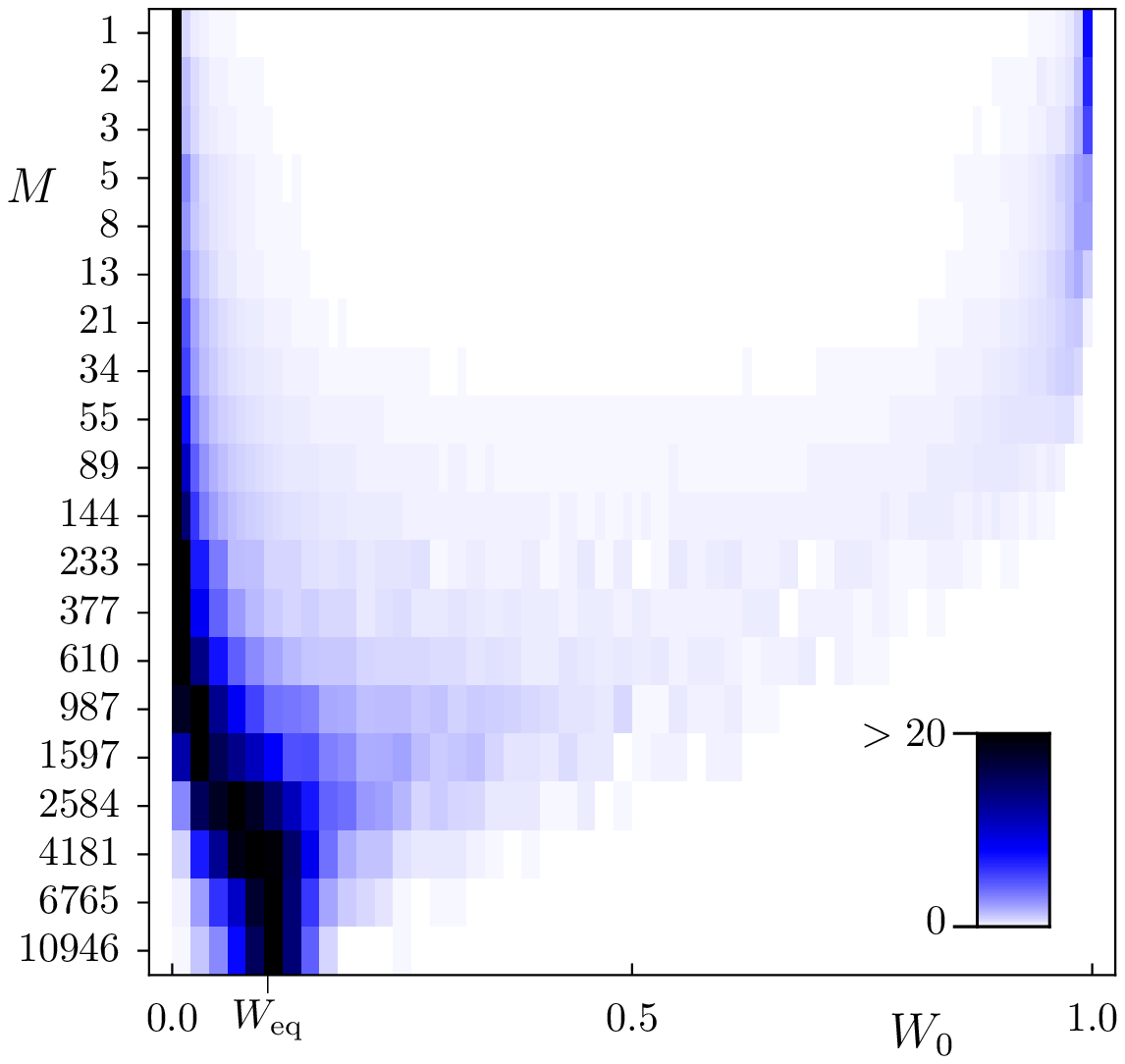}{8.6cm}
    \caption{(color online) 
              Distribution of $\Wnull$ (Eq.~\eqref{weightsemiclass})
              vs system size $M$ for effective Planck's constant
              $\heff \approx 1/10$.
    \label{fig:Weights_m0} } 
  \end{center}
\end{figure}

\begin{figure}[t]
  \begin{center}
    \PSImagx{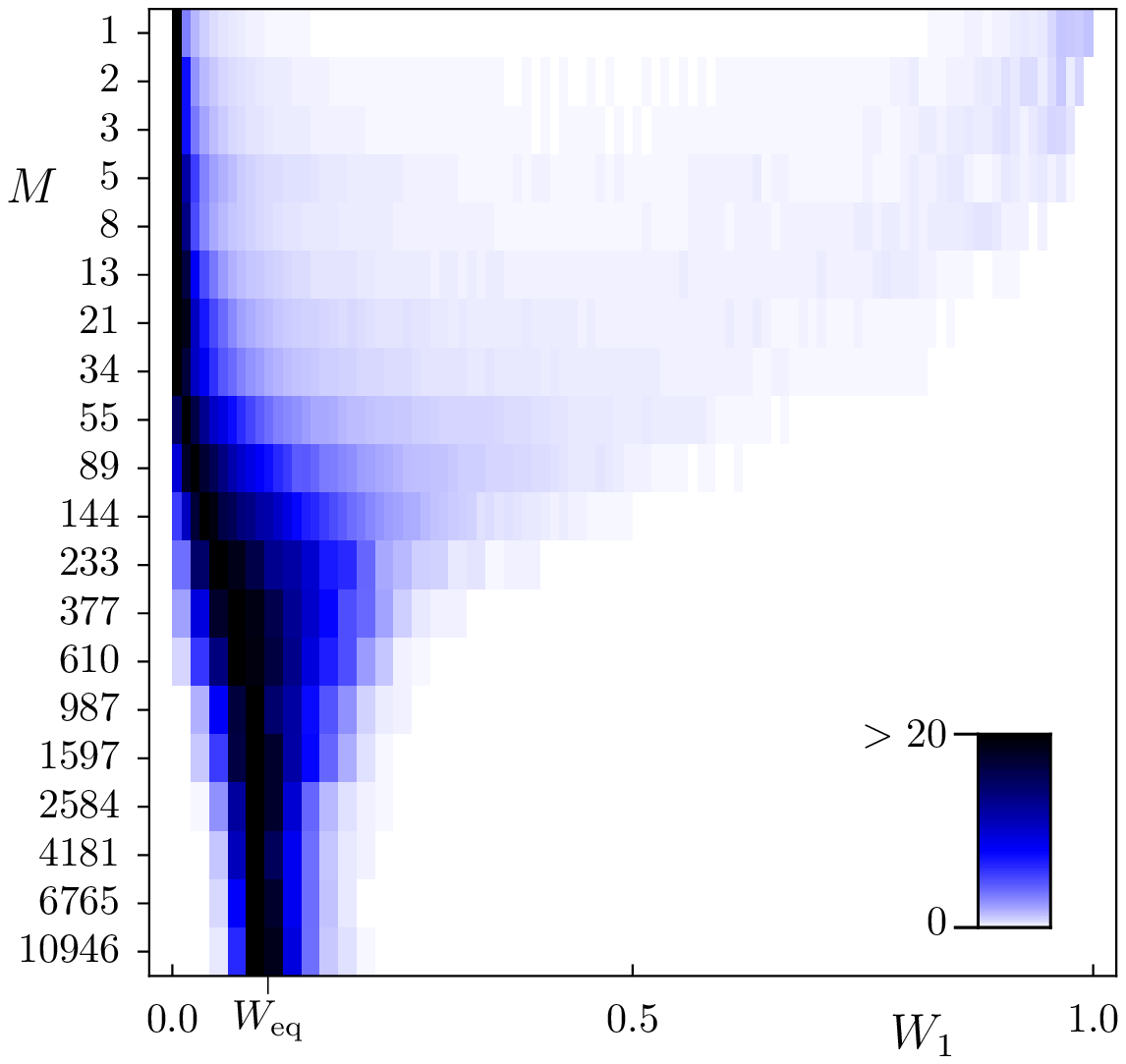}{8.6cm}
    \caption{(color online) 
             Distribution of $\Weins$ (Eq.~\eqref{weightsemiclass})
              vs system size $M$ for effective Planck's constant
              $\heff \approx 1/10$.
    \label{fig:Weights_m1} } 
  \end{center}
\end{figure}

By means of the weight $\Wm$ for all eigenstates of
Eq.~\eqref{Umatrix} we can study the process of flooding for each
torus separately. This allows for a detailed analysis and a
quantitative comparison with a random matrix model. Therefore this is
a considerable improvement compared to our previous analysis
\cite{BaeKetMon2005}, where the weight was defined as the integral of
the Husimi distribution of an eigenstate over the whole region
of the island, which means that the information on individual 
tori is not accessible.

In Fig.~\ref{fig:Avoided}(b) we show the weights $\Wnull$ and $\Weins$
of all the eigenstates as a function of $\chi_q$. For $\Wnull$ we
observe that for almost all $\chi_q$ the weights are essentially zero
or one. Only at avoided crossings of regular and chaotic eigenstates
their weights have intermediate values. For $m=1$ the avoided
crossings are much broader due to the larger coupling and the value
$W=1$ is not reached between several avoided crossings. This is also
seen in the weight distributions shown to the right in
Fig.~\ref{fig:Avoided}(b), where the two peaks from the chaotic
eigenstates (at $W=1$) and from the regular eigenstates (at $W=0$) are
broader for $m=1$ in comparison with $m=0$. Note, that in the situation
of isolated avoided crossings the involved eigenstates are often
referred to as hybrid states.

The distribution of the weights $\Wm$ allows for studying the process
of flooding in a quantitative way. To violate condition
\eqref{newcondition} we need to increase the Heisenberg time, while keeping
the tunneling rates $\gamma_m$ constant. We can achieve this by
choosing a sequence of rational approximants $M/N$ of $\heff = 1 / (d
+ g)$, with $d\in\N$ and the golden mean $g = (\sqrt{5} - 1)/2 \approx
0.618$. This ensures that, while the system size $M$ is increased, 
$\heff$ is essentially kept at a fixed
value, and therefore the tunneling rates $\gamma_m$ are independent of
$M$. Simultaneously, the dimensionless Heisenberg time $\tH = 1 /
\Dcha$ increases linearly with $M$,
\begin{equation} \label{eq:tH}
  \tH = \Ncha = \left( \frac{1}{\heff} - \mmax \right) M  \ ,
\end{equation}
where we used $\Dcha = 1 / \Ncha$ and $\Ncha = N - \mmax M$
is the number of chaotic states. 
Here $\mmax$ is the maximal number of regular states 
in a single island according to the EBK quantization condition
\eqref{eq:EBK}, 
$\mmax = \left\lfloor \Areg / \heff + 1/2 \right\rfloor$.
As discussed in Ref.~\cite{BaeKetMon2005},
$\tH$ may be bounded, due to 
localization effects: For $M$ larger than the
localization length $\lambda$ the effective mean level spacing $\Dcha
\sim (\lambda \Ncha/M)^{-1}$ leads to 
$\tH \sim \lambda \Ncha/M \approx \lambda \heff$,
where $\lambda$ is measured in multiples of a unit cell and
$\Ncha/M$ is the number of chaotic states per unit cell.
For transporting islands, like in
the model studied here, $\lambda \sim 1/\gamma_0$ is unusually large
\cite{HanOttAnt1984,IomFisZas2002andrefs,HufKetOttSch2002},
leading to a maximal value $\tH \sim \heff/\gamma_0$.

In Figs.~\ref{fig:Weights_m0} and \ref{fig:Weights_m1} we show the
distribution of $\Wnull$ and $\Weins$ for $d=9$ (giving approximants
$\heff = 1/10$, $2/19$, $3/29$, $5/48$, $\ldots$) 
for increasing system size
$M$. For small system sizes we increased the statistics by varying the
phase $\chi_q$ in the quantization, as it was shown in
Fig.~\ref{fig:Avoided}(b). To present the results in a compact form
each histogram is shown using a color scale. The horizontal strips
for $M=1$ in Fig.~\ref{fig:Weights_m0} and Fig.~\ref{fig:Weights_m1}
correspond to the histograms previously shown in Fig.~\ref{fig:Avoided}(b).

In Fig.~\ref{fig:Weights_m0} one clearly observes for small $M$
two separate peaks corresponding to chaotic eigenstates 
at $W=0$ \cite{ftn}
and regular eigenstates with $m=0$ at $W=1$. With increasing system
size these regular eigenstates disappear while the weight $\Wnull$ of
the chaotic eigenstates starts to increase 
and they turn into flooding eigenstates.

Comparing Fig.~\ref{fig:Weights_m1} for $\Weins$ with
Fig.~\ref{fig:Weights_m0} for $\Wnull$ one observes a qualitatively
similar behavior. The difference is that the regular eigenstates with
$m=1$ disappear for much smaller system size $M \approx 100$ than the
eigenstates with $m=0$, as expected from Eq.~\eqref{newcondition} and their
ratio of tunneling rates, $\gamma_0 / \gamma_1 \ll 1$.

For the largest values of $M$ only flooding eigenstates are left which
fully extend over the chaotic sea and the regular island. The flooding
is complete and the $N$ eigenstates are equally distributed in the
Hilbert space. Projecting them onto the $M$ regular basis states leads
to the average value $\Weq = M/N = \heff \approx 1/10$, in agreement
with the observed position of the peaks in Figs.~\ref{fig:Weights_m0}
and \ref{fig:Weights_m1} and the findings in
Ref.~\cite{HufKetOttSch2002}.

\section{Random Matrix Model} \label{RMModel}

\begin{figure}[t]
  \begin{center}
    \PSImagx{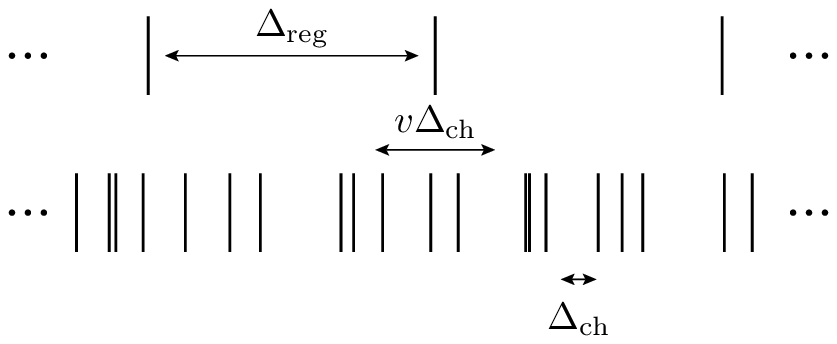}{8.6cm}
    \caption{Schematical plot of the regular levels with spacing $\Dreg$
             coupled with strength $v\Dcha$ to the COE distributed
             chaotic levels with mean spacing $\Dcha$.
    \label{fig:ladder} } 
  \end{center}
\end{figure}

The similarity in the behavior of the histograms in
Fig.~\ref{fig:Weights_m0} and Fig.~\ref{fig:Weights_m1}, 
suggests a universality in the
process of flooding, which should allow for a random matrix modelling.
Such models have been used  successfully for the case of mixed systems
to describe the level splitting in the context of chaos assisted 
tunneling, see e.g.~\cite{BohTomUll93,TomUll1994,LeyUll96,ZakDelBuc98}. 
In our case, we want to describe the statistics of 
eigenvectors for the situation of a chain of $\Nreg$ regular islands.
Here one has $\Nreg$ equispaced
regular levels corresponding to the $m$-th quantized torus and $\Ncha$
COE distributed chaotic levels coupled by dynamical tunneling, see
Fig.~\ref{fig:ladder}. For this situation we propose a random matrix
model with the following block structure
\newcommand{\ENTRY}[1]{\raisebox{0ex}[5mm][3mm]{\makebox[7ex]{$#1$}}}
\begin{equation}  \label{matrix}
H =
\left(
\begin{array}{c|c}
\ENTRY{H_{\rm reg}} & \ENTRY{V}           \\[0.2ex]\hline
\ENTRY{V^T}         & \ENTRY{H_{\rm ch}}
\end{array}
\right) \ .
\end{equation}
This matrix is chosen to be real symmetric because the kicked system 
under consideration obeys time reversal symmetry.
As a consequence of the block structure, the free parameters
of this model are the ratio $\Nreg/\Ncha$ of the
number of regular and chaotic basis states and the strength $v$ 
of the coupling.

The first block $H_{\rm reg}$ models the regular basis states
associated with one specific torus, while for simplicity we neglect
the regular basis states quantized on other tori. As discussed at the
end of Sec.~\ref{TheSystem}~B, in the considered kicked system, the
EBK eigenphases of the $\Nreg$  regular basis
states are equispaced. To mimic this behavior we consider for $H_{\rm
reg}$ a diagonal matrix with elements $(k+\chi)/\Nreg$, $k=0, 1,
\ldots, \Nreg -1$. The parameter $\chi$ can be chosen from a uniform
distribution between zero and one. The energies lie in the interval [0,1] with
fixed spacing $\Dreg = 1 / \Nreg$.

\begin{figure}[b]
  \begin{center}
    \PSImagx{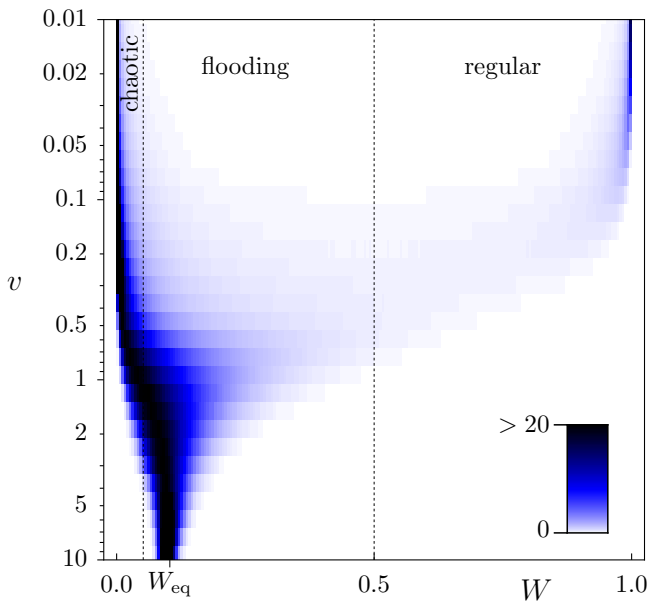}{8.6cm}
    \caption{(color online) Distribution of weights in the random
              matrix model vs coupling strength $v$. The ratio
              $\Nreg/\Ncha$ approximates the value $1/(8+g)$.  The
              dashed lines at $W=0.5$ and $W = 0.5 \Weq
              \approx 0.052$ separate 
              chaotic, flooding, and regular eigenstates.
    \label{fig:Weights_RMT} } 
  \end{center}
\end{figure}

The block $H_{\rm ch}$ models the $\Ncha$ chaotic basis states, where
we assume $\Ncha>\Nreg$. It is also a diagonal matrix whose elements
$\{ E_l \}$ are the eigenphases of an $\Ncha \times \Ncha$ matrix of
the Circular Orthogonal Ensemble (COE). These energies $\{ E_l \}$ lie
in the interval [0,1] with a uniform average density and show the
typical level repulsion of chaotic systems. The mean level spacing of
these basis states is $\Dcha = 1 / \Ncha$. Note, that a GOE matrix for
this block would have been less convenient as it leads to a
non-uniform density of levels according to Wigner's semicircle law.

The off-diagonal block $V$ accounts for the coupling between the
regular and chaotic basis states. It is a $\Nreg \times \Ncha$
rectangular matrix, where each element is a random Gaussian variable
with zero mean and variance $(v\Dcha)^2$. The positive parameter $v$
is the coupling strength in units of the chaotic mean level spacing
$\Dcha$. Thereby the results become asymptotically independent of the
dimension $\Ntot = \Nreg + \Ncha$ of the matrix for fixed $v$ and
$\Nreg/\Ncha$.

We identify the regular region with the subspace spanned by the first
$\Nreg$ components. Therefore, for any normalized vector $(\Psi_0,
\ldots, \Psi_{\Nreg-1}, \Psi_{\Nreg}, \ldots, \Psi_{\Ntot-1})$ 
we define the weight $W$ inside the regular region as
\begin{equation}
W = \sum_{j=0}^{\Nreg-1} | \Psi_j |^2 \ . \label{weightrmm}
\end{equation}
For a particular realization of the ensemble through the numbers
$\{E_l \}$, $\chi$, and the block $V$, we compute the weights $W$ of
the eigenvectors. We take for the statistics only those eigenvectors
whose eigenenergies are in the interval [0.1, 0.9] to avoid possible
border effects.  We determine the distribution of $W$ by averaging
over many different realizations. Increasing the matrix size $\Ntot$
for a fixed ratio $\Nreg/\Ncha$ we find that the distribution
converges. Considering a ratio $\Nreg/\Ncha = 1/(8+g)$ and a small
coupling strength $v \approx 0.1$ the distribution converges around
$\Ntot = 200$. For $v \approx 1$ bigger matrices of $\Ntot
\approx 1000$ are necessary. For $v \approx 10$, we used $\Ntot
\approx 10000$. The limiting distributions depend sensitively on the
coupling strength $v$.

In Fig.~\ref{fig:Weights_RMT} we plot the distribution of $W$ for
different values of $v$.  We have to distinguish between the uncoupled
regular and chaotic basis states of our model and the resulting
eigenstates in the presence of the coupling. The eigenstates fall into
three classes: a) {\it Regular eigenstates} ($W>0.5$), which
predominantly live in the regular subspace. The remaining states,
which predominantly live in the chaotic subspace, are divided into two
classes, depending on the strength of their projection onto the regular
subspace compared to the equilibrium value $\Weq = \Nreg /
\Ntot$. This leads to b) {\it flooding eigenstates} ($0.5 \Weq < W <
0.5$), and c) {\it chaotic eigenstates} ($W<0.5 \Weq$).
Note, that the constants 0.5 in these definitions  are arbitrary.

From the energy scales in the random matrix model, see
Fig.~\ref{fig:ladder}, we expect three qualitatively different
situations for the distribution of $W$:

i) $v\ll 1$, regular and chaotic eigenstates: In this regime the
regular and chaotic blocks are practically decoupled as the coupling
$v\Dcha$ is much smaller than the mean spacing of the chaotic basis
states, $v\Dcha \ll \Dcha$. Two sharp peaks are observable, one at
$W\approx 0$ due to the chaotic eigenstates,
and the other at $W \approx 1$ due to the regular eigenstates. The
latter peak has a smaller weight as the density of regular
basis states is smaller.

ii) $v \approx 1$, chaotic and flooding eigenstates: Here the coupling
$v\Dcha$ is approximately of the same order as the mean chaotic
spacing $\Dcha$.  All regular basis states are strongly coupled to
several chaotic basis states and none of the eigenstates is
predominantly regular.  On the other hand one has different
types of eigenstates as $v\Dcha < \Dreg$: Chaotic basis states, which
are close in energy to a regular basis state, strongly couple and thus
turn into flooding eigenstates. In contrast, there are many chaotic
basis states which are far away from any regular basis state and only
couple weakly. These lead to chaotic eigenstates which show
essentially no flooding ($W<0.5 \Weq$).

iii) $v \gg \Ncha/\Nreg$, flooding eigenstates: All chaotic basis
states are strongly coupled to the regular basis states, $v\Dcha \gg
\Dreg$. The resulting eigenstates equally flood the regular
subspace. The distribution of $W$ gets a Gaussian shape with mean
value $\Weq = \Nreg/\Ntot$ and a decreasing width.

In the transition from situation i) to ii) the two peaks of $P(W)$
near $W=0$ and $W=1$ broaden and move to the center. The regular peak
broadens faster, and at $v \approx 0.25$ its maximum disappears. At $v
\approx 1$ practically no eigenstates are localized in the regular
subspace. When moving from situation ii) to iii) the different types
of chaotic and flooding eigenstates transform into a single type of
flooding eigenstates with a similar weight $W = \Weq$ in the regular
subspace.

How do the resulting distributions depend on the ratio $\Nreg/\Ncha$?
First, the average of $P(W)$ is given by $\Weq = \Nreg/\Ntot
= 1/(1+\Ncha/\Nreg)$. Secondly, the regular peak in situation i) is independent
of $\Nreg/\Ncha$ apart from a trivial scaling of the normalization
with $\Nreg/\Ncha$. Numerically we checked that this is even true up to
$v\approx 1$ for the distribution with $W>0.5$ and $\Nreg/\Ncha \leq
1/(8+g)$. Decreasing $\Nreg/\Ncha$ enlarges the size of the transition
regime between ii) and iii). In particular, the peak near $W=0$ should stay
there up to larger values of $v$.

\begin{figure*}
  \begin{center}
      \PSImagx{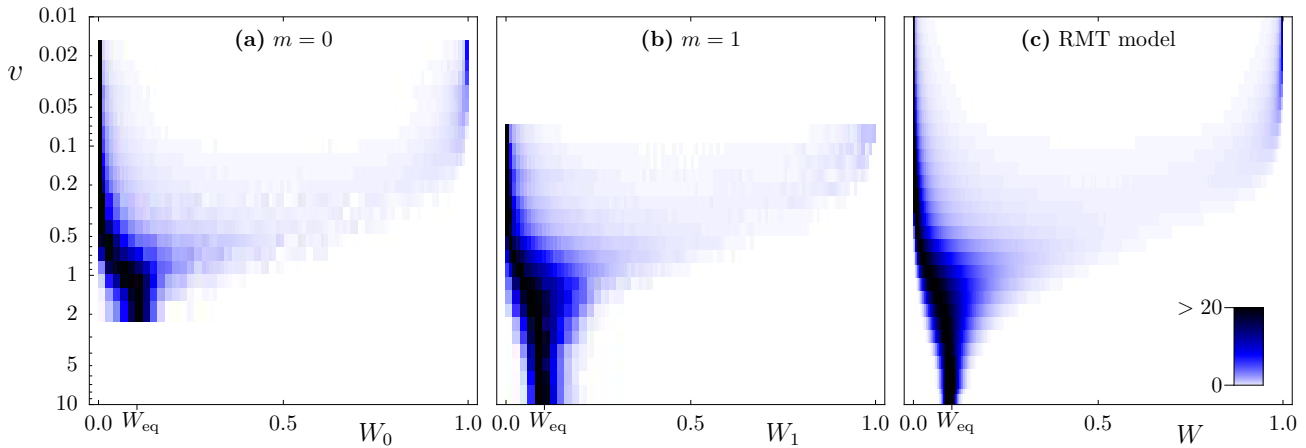}{17.2cm}
      \caption{(color online) 
               Distributions of the weights (a) $\Wnull$ and (b)
                $\Weins$, taken from Figs.~\ref{fig:Weights_m0} and
                \ref{fig:Weights_m1}, with $M$ rescaled to $v$
                according to \eqref{veffectiveB}. (c) Result for the
                random matrix (RMT) model from Fig.~\ref{fig:Weights_RMT}
                on the same scale for a better comparison.
      \label{fig:Weights_RMT_rescaled} } 
  \end{center}
\end{figure*}

\section{Comparison} \label{Comparison}

The distribution of weights for the random matrix model, 
Fig.~\ref{fig:Weights_RMT}, shows a clear similarity to the
results obtained for the kicked system, Figs.~\ref{fig:Weights_m0} and
\ref{fig:Weights_m1}.  In order to obtain a quantitative comparison
one has to determine the relation between the coupling strength $v$ of
the random matrix model and the system size $M$ of the kicked
system. This can be deduced from Fermi's golden rule in dimensionless
form
\begin{equation}
  \gamma = (2 \pi)^2 \frac{\langle V^2  \rangle}{\Delta} \ , 
  \label{fermigoldenrule}
\end{equation}
where the decay rate $\gamma$ of a regular state to a continuum of
states with mean level spacing $\Delta$ is given by the variance of
the coupling matrix elements $V$. In the random matrix model we have
$\langle V^2 \rangle = (v \Dcha)^2$, $\Delta = \Dcha = 1 /\Ncha$, and
therefore \eqref{fermigoldenrule} implies
\begin{equation} \label{veffective}
  v = \frac{\sqrt{\gamma \Ncha}}{2 \pi} \ .  
\end{equation}

Applying this relation to the kicked system, we first note that the
tunnelling rate $\gamma_m$ for each torus can be determined
numerically \cite{Sch:PhD-BaeKetMonSch}
(for recent theoretical results see \cite{OniShuIkeTak2001,BroSchUll2002,PodNar2003,EltSch2005,SchEltUll2005:p,SheFisGuaReb2006}).
The determination of the correct value $\Ncha$ for the kicked system
requires a detailed discussion:
A regular basis state on the $m$-th torus, in the case where the tori
$\mfl, \mfl+1, ..., \mmax-1$ are already flooded, will couple
effectively to $N-\mfl M$ states for $\heff = M/N$. 
A change of $\mfl$ affects
$\Ncha$ and therefore $v$. This dependence, however, can be neglected for
the numerical comparison in our case: The ratio of the maximal and
minimal possible values of $v$ is approximately $\sqrt{ ( 1 - \heff )
/ ( 1 - \Areg)}$. For $\heff \approx 1/10$ and $\Areg = 0.215$ this
gives a difference of less than 7\%. 
Therefore we simply use the
maximal value $\Ncha = N - M$ in the following.

For these values of $\gamma$ and $\Ncha$ in Eq.~(\ref{veffective}) 
the $m$-th torus of the kicked system has a coupling strength
\begin{equation} \label{veffectiveB}
  v = \frac{\sqrt{\gamma_m (1/\heff -1)}}{2 \pi} \sqrt{M} \ .  
\end{equation}
This allows for rescaling the results of the kicked system shown in
Figs.~\ref{fig:Weights_m0} and
\ref{fig:Weights_m1} from $M$ to $v$ using the values $\gamma_0 = 0.0015$
and $\gamma_1 = 0.030$ \cite{Sch:PhD-BaeKetMonSch}. The comparison with
the results from the random matrix model is shown in
Fig.~\ref{fig:Weights_RMT_rescaled}. The agreement is very good for
both tori over a wide range of coupling strengths $v$ showing the
universality of the flooding process. For $v > 5$, however, the
distribution reaches a constant width in
Fig.~\ref{fig:Weights_RMT_rescaled}(b), while the variance decreases
for the random matrix model, Fig.~\ref{fig:Weights_RMT_rescaled}(c).
We attribute this discrepancy to the localization of eigenstates in
the kicked system for $M > 1000$ \cite{HufKetOttSch2002}. 
As a consequence, the effective
number of chaotic basis states near an island saturates
(see the discussion after Eq.~\eqref{eq:tH}),
leading to an effective saturation of $v$.

\begin{figure}[t]
  \begin{center}
      \PSImagx{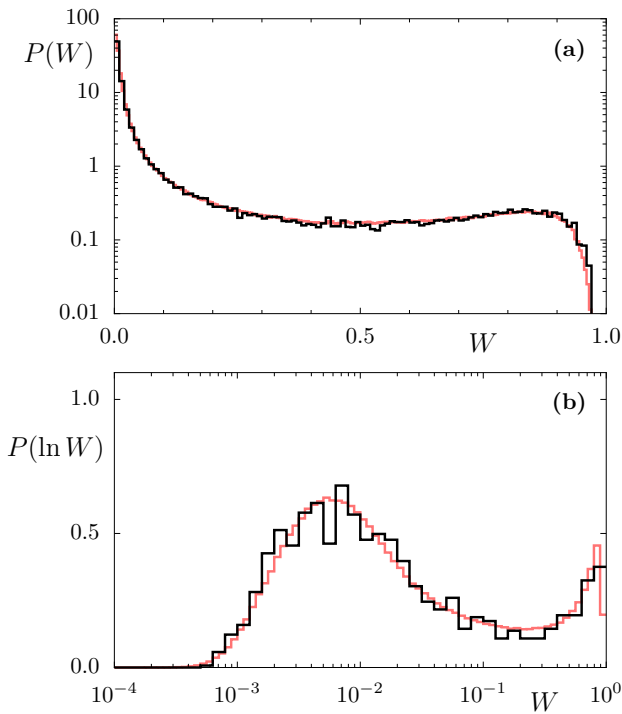}{8.26cm}
      \vspace*{-0.25cm}
      \caption{(color online) 
               Distribution of (a) $\Wnull$ and (b) $\ln \Wnull$
                for $\heff=144/1385$ (dark lines). Results of random
                matrix model for $v = 0.218$ and $\Nreg/\Ncha =
              1/8.618$ (light lines).
      \label{fig:Individual_histogramsA} } 
  \end{center}
\end{figure}

\begin{figure}[h]
  \begin{center}
      \PSImagx{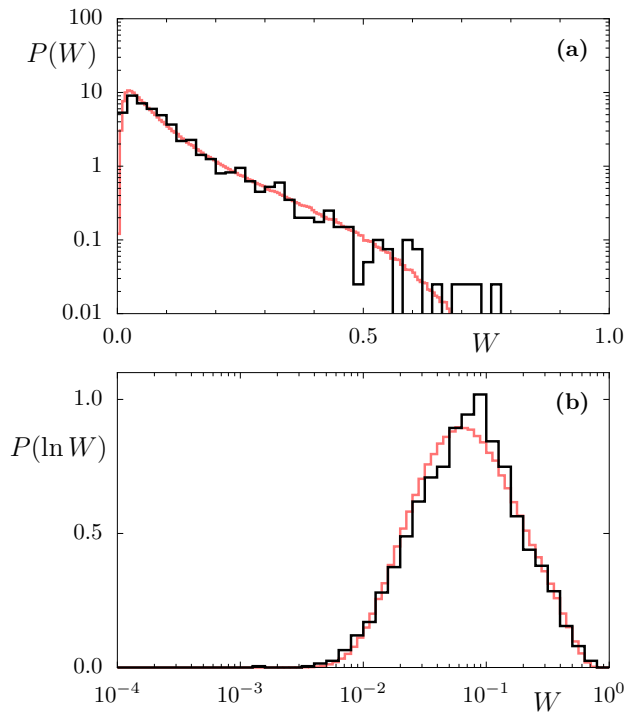}{8.26cm}
      \vspace*{-0.25cm}
      \caption{(color online) 
               Distribution of (a) $\Wnull$ and (b) $\ln \Wnull$
                for $\heff=1597/15360$ (dark lines). Results of random
                matrix model for $v = 0.726$ and $\Nreg/\Ncha =
              1/8.618$ (light lines).
      \label{fig:Individual_histogramsB} } 
  \end{center}
\end{figure}

In Figs.~\ref{fig:Individual_histogramsA} and
\ref{fig:Individual_histogramsB} we compare individual histograms 
for the weights $\Wnull$ for $m=0$. To visualize the low values of the
distributions we choose a logarithmic-linear representation in
Figs.~\ref{fig:Individual_histogramsA}(a) and
\ref{fig:Individual_histogramsB}(a). For $M=144$ one can distinguish the 
peak near $W=0$, due to chaotic eigenstates, from the second peak
caused by regular eigenstates. For $M=1597$ these two peaks have merged
and only a very small fraction of regular eigenstates is left. In both
cases the distributions agree very well with the prediction of the
random matrix model using $v$ according to Eq.~\eqref{veffectiveB}. To
resolve the peak near $W=0$ we show in
Figs.~\ref{fig:Individual_histogramsA}(b) and
\ref{fig:Individual_histogramsB}(b) the distributions of $\ln \Wnull$. 
Again very good agreement with the predictions of the random matrix
model is observed.

\begin{figure}[t]
  \begin{center}
      \PSImagx{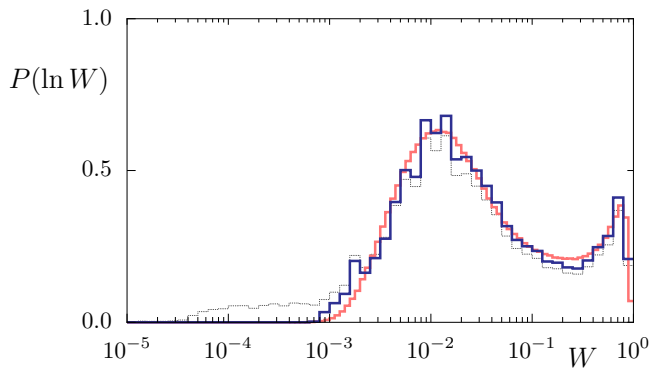}{8.6cm}
      \caption{(color online) 
               Distribution of $\ln \Weins$ of all eigenstates (thin
                line) for $\heff=13/125$. After excluding states
                with $\Wnull > 0.5$ (blue, dark line) much better
                agreement with the random matrix model (red, light line) is
                found.
      \label{fig:Individual_histogramsC} } 
  \end{center}
\end{figure}

Fig.~\ref{fig:Individual_histogramsC} shows the distribution of $\ln
\Weins$ for $m=1$ of all eigenstates for $\heff=13/125$. We observe 
discrepancies at weights smaller than $10^{-3}$ in comparison to the
random matrix model. This difference can be explained as follows:
Among all the eigenstates of the kicked system there are regular
eigenstates localized on the torus $m=0$ which are not considered in
the random matrix model for $m=1$. These eigenstates have a negligible
overlap with the regular basis states with $m=1$ because they are
practically decoupled and only influence the histogram at very small
weights.  This is confirmed by computing the distribution, under
exclusion of all eigenstates with $\Wnull>0.5$.  The resulting
distribution matches remarkably well with the prediction of our
random matrix model.

\section{Fraction of regular eigenstates}\label{Fraction}

\begin{figure}[b]
  \begin{center}
      \PSImagx{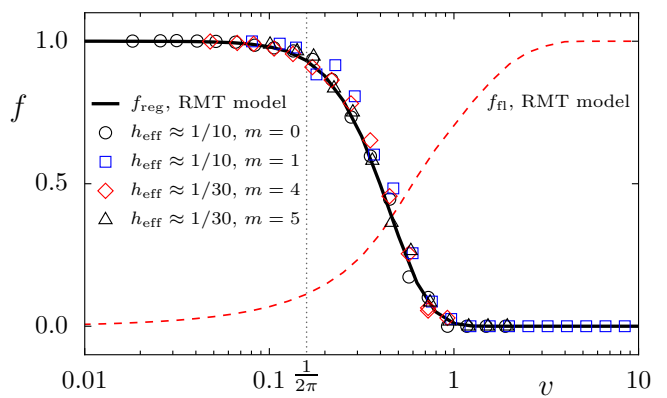}{8.6cm}
      \caption{(color online) 
               Fraction of regular states $\freg$ vs coupling strength $v$ 
                for random
                matrix model (full line) and kicked system for various
                $\heff$ and $m$ (symbols),
                where the system size $M$ is rescaled to $v$ according to 
                Eq.~(\ref{veffectiveB}). 
                Fraction of flooding eigenstates $\ffl(v)$ for the 
                random matrix model (dashed line) for $\Nreg/\Ncha = 1/(8+g)$
                showing a broader transition.
      \label{fig:Fraction_regular} } 
  \end{center}
\end{figure}

A more global quantity than the individual distributions $P(W)$ is the
fraction of regular eigenstates.  This has been studied in 
Ref.~\cite{BaeKetMon2005} for the total number of regular eigenstates as a
function of the system size. With the projection onto individual
regular basis states we are now able to study this fraction for each torus $m$
separately. For the kicked system with $M$ cells there are at most $M$
regular eigenstates localized on the $m$-th torus. However, during the
process of flooding, some of these eigenstates disappear. Thus, we
define the fraction $\freg$ of regular eigenstates on the $m$-th torus as
the number of eigenstates with weight $\Wm > 0.5$ divided by $M$. For
small system sizes this fraction is averaged over several different
phases $\chi_q$. To compare the resulting dependence on $M$ for
different values of $m$ and $\heff$ we determine the coupling strength
$v$ using Eq.~\eqref{veffectiveB}. These results are shown in
Fig.~\ref{fig:Fraction_regular}.

For the random matrix model we compute $\freg$ as the number of
eigenstates with $W > 0.5$ divided by the number of regular basis
states $\Nreg$, averaged over many realizations of the ensemble. As
discussed at the end of section \ref{RMModel}, the distribution $P(W)$
for $W>0.5$ is independent of $\Nreg/\Ncha$, apart from a trivial
rescaling. Therefore the resulting curve $\freg(v)$ is independent of the
ratio $\Nreg/\Ncha$ in contrast to the individual distributions. The
agreement of the fractions determined for the kicked system with the
random matrix curve in Fig.~\ref{fig:Fraction_regular} is very
good. This shows that $\freg(v)$ is a universal curve describing the
disappearance of regular eigenstates. For $v \leq 0.1$ the
fraction of regular eigenstates is larger than 98\%. 
For $v \geq 1$ the fraction of regular eigenstates is less than 1\% 
and the corresponding regular torus is completely flooded.

The criterion \eqref{newcondition} for the existence of a regular
eigenstate, expressed in terms of tunneling rate and Heisenberg time,
can be transformed using Eqs.~\eqref{veffective} and \eqref{eq:tH}, 
into the condition
\begin{equation}
   v < \frac{1}{2\pi} \;\;.
\end{equation}
The position of $v=1/(2\pi)$ 
is indicated in Fig.~\ref{fig:Fraction_regular} and roughly
corresponds to 93\% 
of regular eigenstates still existing (by the $W >
0.5$ criterion). While in Ref.~\cite{BaeKetMon2005} condition
\eqref{newcondition} for the existence of regular eigenstates
was obtained from a scaling argument which does not provide a
prefactor, our random matrix model analysis shows that it is 
quite close to 1.

For the transition regime $1/2 \pi < v < 1$ this model shows 
a decreasing probability for the existence of a regular eigenstate.
For $v >1$, which implies
\begin{equation}
  \gamma_m > (2 \pi)^2 \frac{1}{\tH} , \label{condition_no}
\end{equation}
we find that almost no regular eigenstate exists on the $m$-th torus. 
Thus $v=1$ defines a critical system 
size $M_m$ associated with each quantized torus
\begin{equation} \label{MaximalM}
  M_m = \frac{ 4 \pi^2 \heff}{\gamma_m (1-\heff)} \ .
\end{equation}

\begin{figure}[t]
  \begin{center}
      \PSImagx{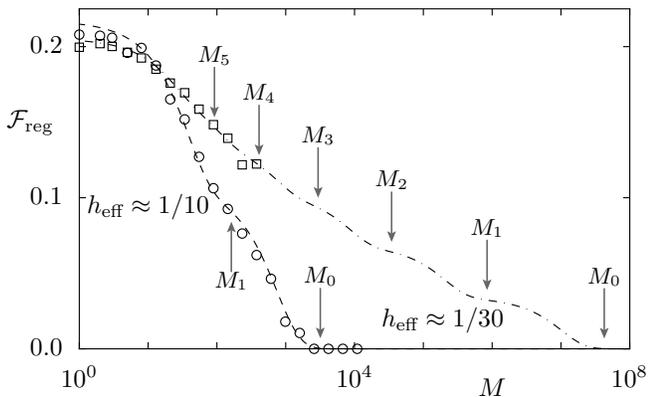}{8.6cm}
      \caption{ 
               Total fraction of regular states $\cFreg$ vs system size $M$
               according to the prediction Eq.~\eqref{TotalFraction},
               (lines) in agreement with the data for the kicked
               system for $\heff \approx 1/10$ (circles) and $\heff
               \approx 1/30$ (squares). 
               The arrows indicate the critical system 
               sizes $M_m$ according to Eq.~\eqref{MaximalM}.
      \label{fig:Fraction_regular2} }
  \end{center}
\end{figure}

With the knowledge about the flooding of individual tori we can
now consider the total fraction of regular eigenstates. The regular
tori with larger $m$ have typically a larger tunneling rate, $\gamma_0
\ll \gamma_1 \ll \ldots \ll \gamma_{\mmax-1}$. Therefore the flooding of the
regular tori happens sequentially from the outside of the island as
the system size increases, as found in \cite{BaeKetMon2005}. The total
fraction of regular eigenstates $\cFreg$ is defined as the number of
eigenstates with weights $\Wm > 0.5$ for any $m$, divided by the total
number of eigenstates $N$. With Eq.~\eqref{MaximalM} we get the
prediction
\begin{eqnarray} \label{TotalFraction}
  \cFreg (M) &=& 
  \frac{M}{N}
  \sum_{m=0}^{ \mmax-1} \freg \left(
  \sqrt{\frac{M}{M_m}} \ \right) \ , 
\end{eqnarray}
where $\freg(v)$ is the universal curve from the random matrix model. 
For small system sizes $M < M_m$ for all $m$ the total fraction of regular
eigenstates is $\cFreg (M) = M \mmax / N \approx \Areg$, as 
expected from the semiclassical eigenfunction hypothesis.
Fig.~\ref{fig:Fraction_regular2} shows $\cFreg (M)$ with a
succession of plateaus and drops before each critical size
$M_m$. Considering that the ratio of successive $M_m$ only varies
moderately, the overall behavior of $\cFreg$ is an approximately
linear decrease on a logarithmic scale in $M$, explaining the
observations of Ref.~\cite{BaeKetMon2005}. The agreement of
Eq.~\eqref{TotalFraction} with the fraction of regular eigenstates for
the kicked system for different $\heff$ as seen in
Fig.~\ref{fig:Fraction_regular2} is remarkably good.

We conclude this section with the remark that due to the independence
of $\freg(v)$ on the ratio $\Nreg/\Ncha$ one can obtain this universal
curve by considering a simpler random matrix model, where only one
regular basis state is coupled to an infinite number of chaotic basis
states \cite{LeyUll96}. For this simpler model it might be possible to
obtain analytical expressions for $\freg(v)$.

\section{Fraction of flooding eigenstates}\label{Fraction-flood}

The random matrix model also allows for investigating the fraction of
flooding eigenstates.  While the regular eigenstates disappear with
increasing coupling strength $v$, more eigenstates turn into flooding
eigenstates with $0.5 \Weq < W < 0.5$.
Fig.~\ref{fig:Fraction_regular} shows  the increasing fraction of these states 
for the random matrix model with $\Nreg / \Ncha = 1/(8+g)$.
Note, that this fraction is defined as the number
of flooding eigenstates divided by the number $\Ntot$
of all eigenstates.  
At $v=1$ all regular
eigenstates have disappeared, however, the fraction of flooding
eigenstates is just 70\%. 
The remaining eigenstates are chaotic, 
which have no substantial weight in the regular subspace.
For larger values of $v$ they turn into flooding eigenstates.
This roughly happens when each chaotic basis state is coupled to at
least one regular basis state, i.e.\ when $v\Dcha =\Dreg/2$, see
Fig.~\ref{fig:ladder}. This gives $v=\Ncha/(2\Nreg)
\approx 4.8$ which is in good agreement with the saturation observed in
Fig.~\ref{fig:Fraction_regular}.
This shows that the fraction of flooding eigenstates explicitly depends 
on the parameter $\Nreg/\Ncha$
in contrast to the fraction of regular states $\freg(v)$.

Applying this result of the random matrix model 
to the kicked system 
where $v=\Ncha/(2\Nreg)\approx N/(2M)$, we find
using Eqs.~\eqref{veffective} and \eqref{eq:tH},
that the fraction of flooding eigenstates is saturated
at $\ffl=1$ for
\begin{equation} \label{eq:all-flooding-states}
  \gamma_m > \left( \frac{\pi}{\heff} \right)^2 \frac{1}{\tH}
\end{equation}
Note, that this prefactor increases in the semiclassial limit
leading to a broader transition to flooding eigenstates.

\section{Summary and discussion} \label{Conclusions}

We provide a detailed quantitative description of the flooding of regular
islands. By using
the projection of eigenstates onto regular basis states, which defines
the weights $\Wm$, the process of flooding can be described 
separately for each torus. The distribution
of these weights in the kicked system agrees accurately with the
distribution obtained by the proposed random matrix model. 
This model depends on two parameters only: the coupling
strength $v$ between regular and chaotic basis states and the ratio of
the number of those states $\Nreg/\Ncha$. The connection of this
coupling strength with the parameters of the kicked system 
is given by Eq.~\eqref{veffectiveB}.

From the random matrix model we gain the following general insights into
the flooding of the $m$-th torus in terms of its tunneling rate
$\gamma_m$ and the Heisenberg time $\tH$:

i) $\gamma_m < \frac{1}{\tH}$: All regular eigenstates on the $m$-th
torus exist. None of the eigenstates predominantly extending over the
chaotic region has substantially flooded the $m$-th torus.

ii) $\gamma_m = (2\pi)^2 \frac{1}{\tH}$: No regular eigenstates on the
$m$-th torus exist. Some of the eigenstates predominantly
extending over the chaotic region have substantially flooded the
$m$-th torus.

iii) $\gamma_m > \left( \frac{\pi}{\heff} \right)^2 \frac{1}{\tH}$:
All of the
eigenstates predominantly extending over the chaotic region have
substantially flooded the $m$-th torus.

What do these results imply for the applicability of
the semiclassical eigenfunction hypothesis?
For a fixed system size in the 
semiclassical limit $\heff\rightarrow0$, 
which implies a roughly exponential decrease of $\gamma_m$, one
ends up in regime i), in agreement with the semiclassical
eigenfunction hypothesis. In contrast,
for small $\heff\neq0$ fixed and systems with $M$ cells and $M\to\infty$,
one obtains a large value for $\tH\propto M$, limited by dynamical localization
only.
Depending on the localization length 
one ends up in regime iii) for some or all tori $m$.
As in our case one has $\tH \sim \heff/\gamma_0$,
regime iii) is realized for all tori, i.e.\ 
complete flooding of the island \cite{HufKetOttSch2002}.

The universality in the transition from i) to ii) can be seen for the
fraction of regular states $\freg(v)$ localized on a given torus. 
For the random matrix model this fraction does not depend on the ratio
$\Nreg/\Ncha$
and the agreement with the results for the kicked system is
remarkably good for different quantized tori and values of
$\heff$. 
In contrast to the disappearance of regular eigenstates on the
$m$-th torus, the transition to flooding eigenstates on this torus is
much broader and extends to regime iii).

It is also important to discuss, what these results imply
for the case of a single island in a chaotic sea ($M=1$).
Most commonly one is in regime i), i.e.\ 
regular and chaotic eigenstates exist and only mix at accidental
avoided crossings.
For a sufficiently small island, compared to the size
of the chaotic region, regime ii) can be reached.
Here $\heff$ is small enough to quantum mechanically resolve
the small regular island, but 
a corresponding regular state does not exist.
It is not possible, however, to get into regime iii) where
all eigenstates would be flooding eigenstates: 
In Eq.~\eqref{eq:all-flooding-states}
we have $\heff = 1/N$ and $\tH =\Ncha\approx N$
such that the right hand side is
approximately $\pi^2 N$, which is always larger
than the tunneling rates $\gamma_m<1$.

In the case of an island chain of period $p$ 
embedded in a chaotic sea it might be possible to
get into regime iii):
In the derivation of Eq.~\eqref{eq:all-flooding-states}
we now have to use $v=\Ncha/(2\Nreg)\approx N/(2p) = 1/(2p\heff)$,
leading with Eqs.~\eqref{veffective} and $\Ncha \approx N = 1/\heff$
to $\gamma > \pi^2/(p^2\heff)$.
The right hand side can be smaller than 1 if $p$ is
sufficiently large while $\heff$ is small enough
to resolve the individual islands of the chain.
Whether this is indeed possible in typical systems 
requires further investigations.

This discussion shows that the semiclassical limit in generic
systems with a mixed phase space, where
islands of arbitrarily small size exist, is rather complicated.
For example one can ask how small does $\heff$ have to be such
that at least one regular state exists on a small island
of size $\Areg$?
Let us define the ratio $r=\heff/\Areg$
The quantization condition Eq.~\eqref{eq:EBK} implies
that $r < 2$ is necessary to quantum mechanically
resolve the island.
However, we find that the necessary ratio $r$ becomes
arbitrarily small for small islands:
Regime i) for $m=0$ requires $\gamma_0<1/\tH \approx \heff$. The
tunneling rate $\gamma_0$ is an approximately exponentially decreasing 
function $\gamma_0 \sim \exp(-C/r)$ with $C$ of the order
of 1 \cite{HanOttAnt1984,FeiBaeKetRotHucBur2006}.
Thus we have to fulfill 
$\exp(-C/r)/r  < \Areg$,
which for decreasing $\Areg$ is only possible if $r$ 
is sufficiently small.

We conclude by emphasizing that the universality given by the random
matrix model not only holds for the kicked system studied here,
but is applicable to any system with a mixed phase space.
The consequences for the semiclassical limit in the
hierarchical phase--space structure of generic systems 
needs much further investigation.

\vspace{0.25cm}
\noindent
{\bf Acknowledgements}
\vspace{0.25cm}

\noindent
We thank S.~Tomsovic, D.~Ullmo and
H.~Weidenm\"uller for useful discussions, Lars Schilling for
providing us with the data for the tunneling rates, and the Deutsche
Forschungsgemeinschaft for support under contract KE 537/3-2.


\begin{thebibliography}{10}
\bibitem{ftn2}
Whether these subspaces are orthogonal or only approximately orthogonal has no
  consequences for the present work and for simplicity we assume orthogonality,
  as e.g.\ in Ref.~\cite{BohTomUll93}.
\bibitem{ftn}
Note, that also regular states on the $m=1$ torus contribute to this peak, see
  Fig.~\ref{fig:Individual_histogramsC}.
\end{thebibliography}
\end{document}